# Assessing the adoption of Virtual Learning Environments in Primary Schools.

## An activity oriented study of teacher's acceptance

Elena Codreanu[1,2,3,4], Christine Michel [1,3], Marc-Eric Bobillier-Chaumon [1,2], Olivier Vigneau[4]

[1] Université de Lyon, Lyon, France
[2] Université Lyon2, GRePS, EA 4163, Bron, France
[3] INSA-Lyon, LIRIS, UMR5205, F-69621, Villeurbanne, France
[4] WebServices pour l'Education, Paris, France

`{elena.codreanu, marc-eric.bobillier-chaumon}@univ-lyon2.fr,`
`christine.michel@insa-lyon.fr, olivier.vigneau@web-education.net`

**Abstract**: This article studies the conditions of use of a VLE (Virtual Learning Environment) by primary school teachers. It first presents a triangulated model to explore Virtual Learning Environments' adoption in primary schools. The theoretical models cover three approaches: the social acceptance, the practical acceptance and the situated acceptance. The situated acceptance of teachers is studied according to the model by using activity theory and qualitative methods (individual and collective interviews). Our study describes how teachers (8 participants) perceived the role of the VLE in the evolution of their working practices (maintaining, transforming or restricting existent practices), in their relationship with parents and in the follow-up of their students.

## 1 Introduction

The definition of Virtual Learning Environments differs from country to country. In UK, the VLEs were designed mainly as pedagogical and collaborative and lately there were added school management tools. In this view, a VLE is *"learner centred and facilitates the offering of active learning opportunities, including specific tutor guidance, granularity of group working by tutor and learners"* [1]. By contrast, in France, the VLEs were since the beginning designed as a unique access workspace, both for school management and for learning activities. The initially management modules (marks, absences) designed for virtual classrooms served then to design pedagogical applications and collaborative group works. In both British and French systems, VLEs aim to encourage communication and collaborative practices between the members of a school community through tools – such as blogging and a messaging service – and to foster access to information (in regards to homework, for example) through the use of a digital planner.

The last report of OECD (Organization for Economic Co-operation and Development) mentions that technologies are not sufficient to support teaching and instructional purposes. They are simple tools in the hands of teachers and it depends on them to take good use in their activities. Yet, our society is *"not yet good enough at the kind of pedagogies that make the most of technologies (...). Adding 21st century technologies to 20th-century teaching practices will just dilute the effectiveness of teaching"* [2, p. 3]. VLEs serve to carry out diverse activities, are intended for several distinct user groups (teachers, students, parents, and staff), and can be exploited in very different contexts: in the classroom, at home or on the move. This complexity can limit the development of practices and the motivation to use it. In this article we chose to evaluate the factors involved in VLEs adoption in primary schools and to consider two processes: technology acceptance and appropriation. When they explain acceptance, the existing studies focus either on individual factors (like satisfaction, effort expectancy) or practical factors (technological features like ergonomic of the system), or, lately, contextual factors (like history and evolution of professional practices). In this article, we propose to present the main theoretical frameworks in the study of acceptance and to eventually describe a triangulated model to evaluate technology adoption. Then we present a situated study analysing the technology acceptance of teachers and the practices they develop.

## 2  Teachers' VLE Acceptance Studies

Some studies analyse the teachers' attitudes to and beliefs about this type of technology. In their study, Kolias et al. [3] examined attitudes and beliefs of teachers from Finland, Greece, Italy and the Netherlands after a first teaching experience with a computer learning environment in order to see if they would be able to include technology in their everyday practices. The study gives very promising conclusion about the possible use of technology, but miss of real practice and acceptance observations.

Others studies analyse the teachers practices and the problems linked with the VLE uses. Indeed, the VLEs have been mainly used in secondary education and higher education. French studies showed that certain teachers had partly integrated VLEs in their professional practices. Prieur and Steck [4] indicated that, although teachers recognized the pedagogical benefits of VLEs, they were not ready to endorse them due to poor ergonomics, and to their lack of training and proficiency in IT tools. Teachers also felt overworked and resisted the idea of extending the *"school space-time continuum"* outside of school. For their part, Poyet and Genevois [5] identified differences in culture: since VLEs are often seen as management tools for businesses, they may need to be "translated" and the meaning adapted to the context of school. One of the ways to solve this issue would be to use school-based metaphors ("notebooks", "lockers") instead of bureaucratic terms ("messaging", "agenda"). Poyet and Genevois showed how VLE tools were unfamiliar to teachers and how the latter did not fully grasp their pedagogical uses and benefits. This led to unsatisfying experimental phases in which teachers tested the tool's various functions, *"without always having a full representation of the tool's potentialities and specific limits"*. This drew teachers to prefer using personal and familiar tools (such as their own emails). Similar observations were made by Pacurar and Abbas [6] who noticed that the VLE was perceived as a communication tool (through the messaging service) and an administrative tool (assigning grades, writing down absences), but that it *"was not firmly anchored in pedagogical practices"*, especially when it came to using it during class time or to design class material. The prescribed uses did not answer the real needs felt by teachers on a daily basis. These conclusions are also given by Firmin and Genesi, [7] and Blin and Munro [8]. Bruillard [9] mentioned the complexities in deploying VLEs when a variety of people are involved: teachers, parents, students, school districts, local authorities, software publishers and the Ministry of Education. Bruillard also noticed a paradox between the Ministry's will to open schools up to parents, and the actual low amount of parental implication. Teachers are also concerned that parents may interfere in their pedagogical choices. These difficulties are further amplified by the fact that teachers who use VLEs do not get institutional recognition. Practitioners in the field have also felt disempowered since external companies were called to design the VLEs. There is also the risk of creating inequalities or even to exclude certain parents who are less equipped and trained in digital technologies. Missonier [10] developed these points based on the design and the deployment of VLE projects that were managed by local authorities and service providers. These approaches have not always been very effective, since they depend on the project manager – who may lack in transparency or carefulness – to solve disputes linked to functionalities or uses. This, in turn, leads to different protagonists within the network to decrease their commitment. Prieur and Steck [4] recommend implementing spaces for ideas *"that articulate the current practices of teachers, practices that can help foster the acquisition of skills and the potentialities of different VLE tools, in order to develop possible instrumentalisations"*. This would help to adapt prescribed uses, depending on the context.

Voulgre [11] introduced a political dimension. Teachers are generally favourable to arguments promoting the uses of VLEs: the latter are useful to catch up on classes (illness, loss of grades), to retrieve previous work or to support students with schooling difficulties. But the fact that not all children have Internet at home represents an inequality, thus preventing teachers from fully using VLEs. Such a refusal is seen as a "*type of counter-power*" against political injunctions. On the contrary, acceptance factors are linked to the respect of hierarchy, of the institution and of the law (obligation to use a VLE); other positive factors are linked to the values of solidarity and cooperation that are promoted by VLE tools.

Other studies also point out the importance of technical infrastructure: access to the computer classroom, number of computers in classrooms, Internet access, broadband speed and technical support. The school institution's management, the organisational culture and VLE implementation strategies have all a great role in technology acceptance [12][13][14][15]. Finally, lack of competences in technology, lack of confidence and lack of time were mentioned [16]. In the end, all of these studies showed that the acceptance of VLEs by teachers depended on practical considerations, as well as strategic concerns that were both professional and political.

VLE began to be deployed now in primary schools. Only a few studies explored the acceptance of VLE in these contexts. Berry [17] highlighted that primary school pupils can use VLEs and appreciate it in case of absence because they can easily get lesson content and homework. Moreover, they have more confidence to discuss mathematics problems on the VLE platform. But younger children differ greatly from stu-dents in secondary or higher education in terms of their autonomy and their use of digital media. So we are led to ask ourselves how primary school teachers take this factor into account and more generally how they include such a new tool in their professional practices: are they able to adapt or develop their practices or not and what are their reasons?

We need to evaluate how actual teaching practices can evolve in order to integrate and make profit of the existing technologies. This is why we aimed in this field study to identify the current teaching practices that constitute the core of professional activities for primary school teachers. We also wanted to identify tensions that could lead us to find ways to improve the design of VLEs and to provide recommendations for uses and services.

## 3 Analyzing Acceptance and Appropriation

### 3.1 The Models of "Social Acceptance"

These approaches focus on human factors in the process of technological acceptance. The main idea is that people's perceptions and attitudes may play a major role in this process. According to Davis [18] and his model TAM (Technology Acceptance Model), acceptance can be explained through two factors: perceived usefulness and perceived ease of use. These two perceptions influence the intentions to use the technology which, in turn, influence its acceptance. Other attitudinal factors are later added: satisfaction, performance expectancy, effort expectancy. This model is inspired by the theory of reasoned action [19] which consider that behaviour is guided from inside by people's intentions. Other authors [20] talk about internal factors (like beliefs, convictions and attitudes of users), and external factors (like support, training, technical infrastructure). Some authors support the idea that internal factors take priority in the decision to use an educational technology [21][22] while others think that external factors are predominant [23]. When they study VLEs acceptance in particular, authors highlight the same duality. While some support the major role of technical infrastructure like access to the computer classroom, number of computers in classroom, Internet access and high speed Internet access and institution management [12][13][14][15], others admit that causes of VLEs reject are lack of confidence in technology and lack of time to train [16]. Other studies show that it is actually the connection between the internal and the external factors that matters: external factor (like institutional support, training) will subsequently shape the beliefs and attitudes toward the technologies and then the intention to use those [24].

In primary teaching, technologies are less frequent, so there are not many studies on this particular subject. Studies demonstrated the importance of self-confidence toward computer use in the development of attitudes toward technologies and indirectly in the intention to use the technologies [25][26][27][28]. Beside confidence, some authors outline the role of perceived security in the acceptance of VLEs in primary school [29]. VLEs suppose a functioning similar to that of social networks, with a unique access to content. Teachers doubt their own possibilities of control and moderation in cases of on-line bullying and interrogate about the responsibilities in case of misappropriation of the VLE by students. Also, they worry about the misuse identity by other colleagues. In primary schools, these issues are particular important, because the students are particular young and vulnerable to these forms of harassment.

Social acceptance approaches have nonetheless been subject to a number of criticisms concerning both methodological criteria and the models' foundations [30]. One criticism is that these studies have little practical relevance for the technological design and improvement of the system. In effect, these studies indicate that a system is not acceptable to the target group without giving any information about the changes and adaptations required. Added to this is the fact that the research is based on small samples that are not representative of the professional context, and use questionnaires (scale of measurement) as the sole method of evaluation. Critics claim that such a method results in a truncated, partial and rather disembodied picture of the meaning people attach to the technology. However, in educational context, we retain the effort to specify precise factors directly implied in technological acceptance: confidence in computer use, social and institutional support, technological infrastructure and children's security.

### 3.2 The Models of "Practical Acceptance"

This approach focuses on the technology characteristics (human factors and ergonomics) and how the tool is implemented (support, training, participatory design). The prevailing idea is that when technology is easy to use and well implemented (training is provided and end users are included in the design process, for example) the device's acceptance is enhanced. In sum, the aim is not only to design a suitable product, but also a suitable relationship to technology, and ultimately contribute to an acceptable user experience for the individual [31].

According to Nielsen [32], the two most important attributes for technology acceptance are usability and utility. Usability refers to the fact that people can easily use the functions of a system. Utility refers to the capacity of the system to help users do their tasks. In short, a technology easy to use and useful will be accepted by the users. To these two attributes, Nielsen adds others: costs of the technology, compatibility, and reliability. We have to mention that the notion of "usability" is different of that of "perceived ease of use" in the previous social model. While the first refers to the effective usability and is evaluated through user tests, the second refers to perceptions and subjective attitude toward usability and is evaluated through questionnaires. The ISO 9241 norm specify that the three dimension of usability are: effectiveness (the accuracy with which users achieve specified goals), efficiency (the effort required for users to do theirs tasks) and satisfaction: what users think about the system.

Ergonomics specialists proposed a list of criteria to evaluate the usability of computer interfaces. Bastien and Scapin [33] proposed eight criteria: guidance (means available to orient the user throughout the interface), workload (interface elements that play a role in the reduction of users' perceptual and cognitive load), explicit control (the control users have on the processing of their actions), adaptability (the system's capacity to behave according to users' needs), error management (means available to recover from errors), consistency (maintaining the interface choices in similar contexts), significance of codes (codes and names should be meaningful for users) and compatibility (match between the users characteristics and task characteristics). Concerning the last criteria, compatibility is particularly important when technologies are used by users with specific characteristics (in terms of age, customs, perceptions, skills). For instance, technologies designed to be used in primary schools, should be adapted to a public of young children, who do not master writing, reading and have limited fine motor skills. So, the interfaces should avoid using a lot of text content and complex pull-down menus; they should prefer instead images and simple menus [34][35]. Budiu and Nielsen [36] used specific methods in order to evaluate children's behaviour on the web (think aloud, card sorting). They proposed a list of 130 recommendations for interfaces designed for children (aged 3 to 12), organised by the type of content (general interaction, navigation, images, videos etc.). Generally, they recommend to use interactive content, sound and colours, use of the metaphors and big buttons. They also advise to ensure children's control over the interface and to avoid sensory and cognitive overload.

These studies are important because they provide precious practical advising for designers. The main criticism is that they are focused on functional aspects and do not consider the intrinsic characteristics of user like emotions (pleasure, fun, amusement). Recently, studies began to consider user as a real partner in design of a technology in approaches like User Centred Design and participatory design [37][38]. Participatory design "relies on the collective generativity of stakeholders; in other words, it uses the collective ability of stakeholders to generate or create thoughts and imaginings" [39, p. 173]. In school technologies, participatory design suppose that teachers and students can be actively involved in the design of their future tools so that these tools would better meet their needs [40][41][42][43].

This approach focuses therefore on the technology conception, on ergonomic improvements and on support to collaboration between designers and end users. In this context, ergonomic approaches intend to prescribe recommendations and guidelines for designers in terms of technological adaption to users 'needs. However, these studies remain focused on the functional aspects and on the performance of users with the system. In addition, participatory design, mostly applied in industry, is less adopted by the stakeholders in digital education. This is due, on one side, to the difficulty and high cost of putting participatory design into practice and, on the other side, to the diversity of educational contexts and high number of schools, with their own autonomy and specific organization which make technological generalization difficult.

### 3.3 Appropriation and Situated Acceptance Models

According to Jonsson [44], appropriation is "the gradual process by which participants successively become more proficient in using the tools" (p. 11) Unlike mastery, which entails the acquisition of a skill, appropriation, in addition to a technical skill, includes the competence to use the technology for carrying out an authentic task in a given context. As such, appropriation is thought to be strongly linked to the notion of change. Using a text editor at school changes practices very little, but being able to modify a digital text without having to copy it out can change the importance traditionally attached to writing. Bobillier-Chaumon [45] considers that the appropriation of a technological tool is a condition of its acceptance. When someone appropriates a tool, she contributes to it and is able to innovate, and therefore use the tool for previously unforeseen purposes. By making this contribution to the technology, the person can identify with it, make it her own, give it meaning and therefore accept it. Bobillier-Chaumon proposes the notion of situated acceptance, defined "as the way in which an individual – or a group or organization – perceives the issues related to these technologies (strengths, benefits, risks, opportunity) through their use in everyday situations, and reacts to them (favourably or not)." [46] What is taken into account here is the experience in a situation of interaction between users and a certain technology that already exists. In this approach, the object of study is not the perception or attitude towards technology but the practices and activities carried out as part of a real job. The advantage of this approach is that it brings into light for the first time dimensions like "history" and "context" and proposes to look for acceptance directly in daily activities of end users.

Activity theory, as detailed by Engeström, Miettinen and Punamaki [47] and Kuutti [48], provides more complete elements to quantify the context of use. Instead of referring to uses, activity theory refers to an activity system: the user (subject) has a precise objective and accomplishes it by using certain instruments (tools). He/she fits into a social community (the group of people who intervene in the activity). This community is regulated by certain operating rules (the norms and rules to respect in a given activity), and respects specific divisions of work (the ways in which roles are distributed among individuals).

Activity systems are characterized by contradictions (or internal tensions), which favour and trigger innovation; such changes contribute to further development. Therefore, activity theory appears to be useful to qualify the context as well as to define the dynamics at work when accepting and taking ownership of technology.

The teacher's activity system is summarized in Figure 1 and relates to the educator's daily practices. These practices occur with or without instruments, since they often take the shape of direct communication in class, and can be supplemented with instruments such as the board, posters, notebooks, etc. These practices follow rules that are specific to the educational system and fit into an educational community composed of teachers, students and parents. The division of work includes the effective practices inherent to the profession and the ways in which the different tasks are distributed among the different protagonists. In terms of the education and follow-up of students, teachers and parents work together, but in different contexts. Each group's responsibility is therefore well defined. With the arrival of a new technological tool, used both in class and at home, these differentiated roles and identities may come into conflict.

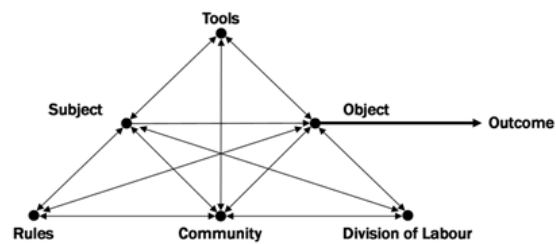

**Fig. 1.** Activity system [47]

Furthermore, according to Rabardel and Bourmaud [49], the conditions needed to implement human-machine interactions lead to the modification of the technology's properties and, consequently, to the readjustment of human conducts. This occurs through the process known by Rabardel and Bourmaud as the instrumental genesis (a double process of instrumentation/ instrumentalisation). The tool therefore does not only exist for itself or in an isolated way. It is socially embedded and fits within certain practices, habits and social communities that guide its use and transform its characteristics.

### 3.4 An Analytical Model of VLE Adoption in Primary Schools

We have identified three categories of approaches. The first, social acceptance, focuses on the individual perceptions and attitudes of prospective users; the second, practical acceptance, concentrates on the tool's ergonomic characteristics; and the third analyses users' activities and hence the interaction between the technology and actual practices. In our study, we need to evaluate the acceptance of a VLE, a complex tool designed for multiple user groups (teachers, students and parents) to perform diverse tasks in a range of contexts (communication, learning, monitoring, etc.). We assume that practical acceptance and situated acceptance impact the social acceptance and perception factors that determine the development of the use and the long term acceptance and appropriation. Consequently, we consider that acceptance is a process that can be evaluated through three sets of factors:

- Technological factors (practical acceptance approaches) grouped in system quality factors (like usability) and design quality factors (participatory design),
- Activity and task factors (appropriation and situated acceptance approaches) related to characteristics of professional activity like rules, prescriptions, professional practices, objectives,
- Perception factors (social acceptance approaches) related to individual opinions about the qualities of the technology (perceived ease of use, satisfaction, perceived security).

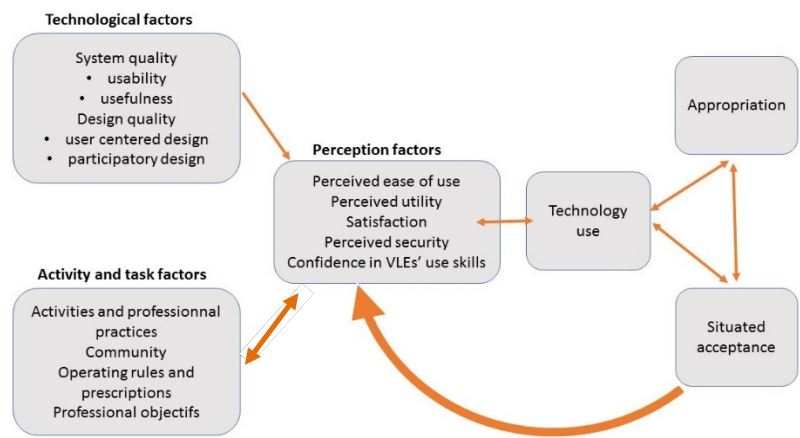

**Fig. 2.** Model of VLE adoption: factors involved in the acceptance and appropriation of VLEs.

In the above diagram (see Figure 2), the single arrow indicates a one-way relationship between the two factor categories. The double arrow indicates a two-way relationship.

The technological factors (quality of the product, quality of support) influence the perceptions of the tool, which in turn influence the tool's appropriation and acceptance. For their part, the activity and task factors (activity, practices, community) also influence the perception factors. The creation of technology's meaning is made during the actual use. The use trials influence significantly the level of technology acceptation and appropriation. The quality of use will build a new form of appropriation (by creating new forms of practices and innovative use) and acceptance (through the lens of new emotions and new benefits related to use). These two constructs will modify the initial perception of the technology and the users' perceptions on their technological skills. The retroactive loop describes how appropriation (seen as mastery of the tool plus innovation) is decisive for the acceptance of the tool (seen as the subjective decision to start using the technology) and vice versa.

It is a dynamic model that may enable the plurality of viewpoints and situations to be reconstructed. Dynamism of the model is important for explaining the principles of technology adoption through articulation of factors issued of different theoretical approaches. This model may restore a diversity of points of view and situations and the formalisation of factors' progression in context. In order to deepen this approach and qualify the criteria of each factor, we propose to use triangulated methods [50] which consists in using more than one method to study a phenomenon. In terms of methodology, we propose a triangulation consisting of qualitative methods (interviews, elicitation interviews, content analysis) and quantitative methods (questionnaires, analysis of connection logs).

## 4   Field Study Methodology

According to this model, we choose to focus the first observations on technological factor, activity and task factor, perception factors and use to answer the questions "How primary school teachers include a new tool like VLE in their professional practices? Are they able to adapt or develop their practices or not and what are their reasons?" We propose to use the activity theory to detect VLE acceptance and non-acceptance factors according to contexts of use. The standards considered to define acceptance are linked to the ways in which the profession is practiced, to social and work constructs and to ways in which the VLE tool is used and deployed. The approach developed in this study is essentially qualitative. We aimed to collect testimonies from teachers in which they represented and perceived their experiences as they teached with and used a VLE.

### 4.1   Observed Context and Participants

All participants in our study were part of the Versailles and Caen school districts (situated near Paris). 6 schools were in the Versailles district and 6 were in the Caen district. They volunteered to experiment with the VLE One for 2 years. At the time of our study, 26 teachers (in both districts) had volunteered to be part of the experiment and had already used the VLE One for 3 to 6 months.

We questioned 8 teachers over the course of 4 individual interviews and 2 group interviews (with 2 teachers in each interview). Among the teachers, two were school principals who were also giving classes (in first and fifth grades). The other teachers worked in first grade (2), second grade (1) and fifth grade (3) classes. The group of participants was composed of seven women and one man. The schools were all situated in urban areas, in the Versailles school district (6) and in the Caen district (2). The average age of participants was of 46 years with a standard deviation of 15.

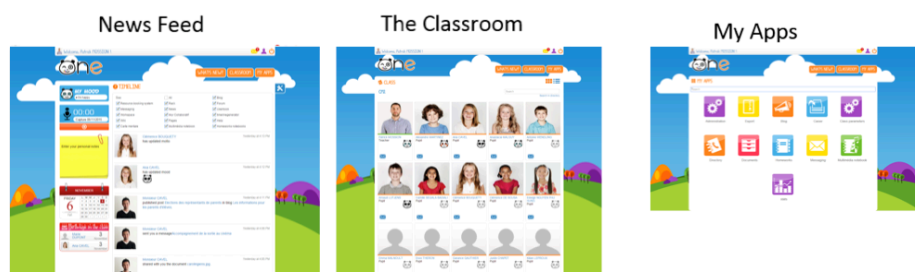

**Fig. 3.** Interfaces for the pages « News Feed » « The Classroom » and « My Apps » in the VLE One

### 4.2 Description of the Tool

The VLE used in this study is entitled One. It was specifically designed for an elementary school audience, with ergonomics and interfaces that are suitable for children [35][36]. The One interface is therefore simple, intuitive and attractive (see Figure 3). The collaboration functions that are offered consist in a Messaging Service, a Blog and a Storage Space. One also offers customization features (My Account, My Mood), notifications (a News Feed, birthday notifications), organizational tools (Calendar) and a school website. Each user has the option of customizing his/her profile with a picture and personal information (motto, mood, information on favourite leisure activities, films, music, food). Students are by default included in their class group and have access to the content published in the group by the teacher.

When we were conducting our study, the VLE One had not yet offered services such as the Planner notebook and the Multimedia notebook.

### 4.3 Data Collection

Teachers participated to semi-structured interviews. These interviews lasted an hour and a half on average and tackled the following themes: the teachers' experience with TEL (Technology Enhanced Learning), the school's computer equipment, the teacher's representation of the VLE, needs related to the VLE, the VLE's usefulness, ease of use and intentions of use, difficulties of use, and the implications of the VLE for the teaching profession.Teachers could speak openly and were able to give their critical point of view on various uses, share their own representations of the tool, and give their opinion on functions that were being developed, such as the planner notebook, the digital parent-teacher notebook and the multimedia notebook. They were also welcome to recount difficulties linked to the use of the VLE, using Flanagan's critical incident technique [51]. Technological, task and perception factors considered in the model can so be described.

### 4.4 Analysing the Teachers' Interviews

The interviews were entirely recorded and transcribed so that they could be systematically studied [52]. We considered in our analysis the comments that associated One with daily teaching practices, operating rules (linked to the educational system), the education community (composed of teachers, students and parents) and the division of work (the ways in which tasks are shared between different groups of people). We used the sentence – a basic syntactic unit built around a verb – as the main unit to study the transcripts. Sentences were identified as in the following example: *"I showed them how to make folders (sentence 1)/, but it is hard for the students (sentence 2)"*. We also distinguished between the comments that were rather favourable (supporting initiatives) and the ones that were less favourable (difficulties in use). We proceeded to do counts and percentage calculations to rank the different factors. We determined that the users had accepted the VLE when they mentioned the successful ways in which they used it, the adjustments they made or the contradictions they encountered and overcame. Categories weren't pre-established and we retained the themes that had been mentioned at least three times.

## 5   Results

The analysis revealed 4 main themes (see Table 1), as well as 16 sub-factors (see Table 2): (1) factors linked to the practice of the profession (the workload, raising awareness of digital uses and habits, work recognition), (2) factors linked to pedagogical monitoring (pedagogy, health and safety, emotions and attractiveness); (3) factors linked to social and work-related organization (collaboration, communication, the reorganization of communicative practices), (4) factors linked to the tool's use and deployment (ease of use, usefulness, feedback, computer and network equipment, support and assistance). We will first present the results that stemmed from the four main factors; we will then proceed to describe the sub-factors.

### 5.1   Main Factors

In Table 1, we can see that the factors linked to social organization brought about the largest number of positive comments (88), which means that the VLE played an important role in communication and collaboration practices within the school activity system. Conversely, factors linked to the teaching profession and to the use and deployment of the VLE gathered the largest number of negative comments. The deployment and use of the VLE therefore seem to raise questions linked to professional recognition and to the practice of the teaching profession. It also raises issues regarding the alignment of VLEs with school uses and habits. In the following paragraph, we present an analysis according to each sub-factor (see Table 2), thus allowing us to refine each element.

**Table 1.** Main Factor Occurrences

| Factor | Number of positive comments | Number of negative comments |
|---|---|---|
| Profession | 35 (15,56%) | 90 (36%) |
| Pedagogical follow-up | 54 (24%) | 57 (22,8%) |
| Social organisation | 88 (39,11%) | 14 (5,6%) |
| The tool's use and deployment | 48 (21,33%) | 89 (35,6%) |
| **Total** | **225 (100%)** | **250 (100%)** |

**Table 2.** Sub-factor Occurrences

| Sub-factor | Number of positive comments | Number of negative comments |
|---|---|---|
| **Factors linked to the practice of the profession** | | |
| Workload | 0 (0%) | 72 (28,8%) |
| Raising awareness on digital uses | 20 (8,89%) | 12 (4,8%) |
| Work recognition | 15 (6,67%) | 15 (6%) |
| **Total** | **35 (15,56%)** | **90 (36%)** |
| **Factors linked to student monitoring** | | |
| Pedagogy | 20 (8,89%) | 0 (0%) |
| Health and safety | 4 (1,78%) | 57 (22,8%) |
| Emotions and attractiveness | 30 (13,3%) | 0 (0%) |
| **Total** | **54 (24%)** | **57 (22,8%)** |
| **Factors linked to social and work-related organization** | | |
| Collaboration | 12 (5,33%) | 0 (0%) |
| Communication | 72 (32%) | 8 (3,2%) |
| Reorganizing communicative practices | 4 (1,78%) | 6 (2,4%) |
| **Total** | **88 (39,11%)** | **14 (5,6%)** |
| **Factors linked to the tool's use and deployment** | | |
| Ease of use | 27 (12%) | 24 (9,6%) |
| Usefulness | 9 (4%) | 6 (2,4%) |
| User feedback | 4 (1,78%) | 39 (15,6%) |
| Computer and network equipment | 0 (0%) | 6 (2,4%) |
| Support and assistance | 8 (3,56%) | 14 (5,6%) |
| **Total** | **48 (21,33%)** | **89 (35,6%)** |

## 5.2 Factors Linked to the Practice of the Profession

As we can see in Table 2, the perceived workload (triggered by the use of the VLE) brought about the largest number of negative comments (72). In fact, teachers had the impression that they needed to invest additional time to master the VLE's functionalities and to imagine interesting projects to do on the platform. They also felt that using the VLE implied sustained and continuous work for new tasks that did not necessarily fit into their areas of expertise, such as: taking pictures, downloading material on the computer and then on the VLE, publishing blog posts, writing messages, and designing teaching projects that included the VLE. Since these teachers did not have a dedicated time slot to use these technologies, they had to use pedagogical time to become familiar with such tools. Teachers also felt the weight of large workloads, with the impression of having an ever increasing amount of informational solicitations. The VLE had indeed been added to a number of pre-existing educational platforms: academic e-mail, the career management platform "I-prof", online training platforms, didactic platforms and an online handbook of skills. Teachers therefore felt constantly submerged by a large amount of data which they had to manage (email addresses, different login names and passwords for each platform, various approaches and functions according to the different resources...). They also felt overwhelmed by the informational content that they had to focus on and prioritize (academic information, pedagogical information, event notifications to sort and share...). Faced with the fear of having to work twice the amount with a VLE, some teachers refused to publish their lessons on the VLE since they already did the same thing using their own automation tools: "*I already create the lesson on "paper board", so putting it up again (on the VLE)... I do not want to do that...*"

Teachers made 20 positive comments about making students more responsible when using digital tools. Teachers found that they had a part to play when training "*students to use digital tools responsibly*". On the other hand, some teachers found that parents should be in charge of raising their children's digital awareness (12 comments). These teachers' main arguments had to do with the fact that working on the students' digital responsibilities affected other teaching activities negatively. They also argued that such digital tools were massively consulted by the children at home, such as when they checked new messages. For these reasons, controlling digital tools should relate to the private sphere. This opinion was not necessarily shared by parents who believed that, on the contrary, the follow-up on digital practices should be done by the institutions that set up the tools in the first place. We can therefore see that, within the "school-home" axis, responsibilities and roles between teachers and parents may need to be redefined within the teaching program, and the division of work would need to be more efficiently coordinated (controlling and following up on uses).

Work recognition was mentioned positively 15 times. Some teachers saw the VLE as a way to highlight classroom work through the blog. Some activities, which had previously been almost invisible to parents, could now be displayed, such as sporting activities, class outings, and the work of the pupils themselves. The VLE then became a tool that could help recognize the teacher's and the students' work. But such recognition is still limited due to parents not being fully involved in the VLE project and not consulting these resources often (negative mentions).

### 5.3 Factors Linked to Student Monitoring

According to the teachers, the primary benefit of VLEs for students lied in the fact that VLEs helped to build a more attractive and stimulating relationship based on emotions (30 positive comments in Table 2). The VLE was a motivating tool for students and allowed them to appreciate class work. In terms of pedagogy, the VLE was seen as a benefit (20 positive comments) in the construction of verbal expression and student communication. It was also positively viewed to raise awareness and autonomy when students were working with computers. The VLE blogs were therefore often co-edited by the teachers and the students.

However, teachers also expressed many fears linked to the children's health and safety (57 negative comments versus 4 positive ones). These fears related more specifically to possible abuses (bullying, insults) or to the misuse of communication and coordination tools. Teachers did not gave any access to the children's accounts and were therefore unable to control the content of exchanged messages. Several teachers created a fictitious student account to follow and control exchanges. This also allowed them to check the layout quality of the information and documents that they published on the VLE. We noticed that the teachers who had not used the platform in such an innovative way weren't as satisfied with the device. This example highlights the importance of offering verification and surveillance functionalities for the teachers, with parent or student views available. Another fear related to ways in which the children themselves could use the VLE in transgressive ways. It is particularly difficult for teachers to authenticate information coming from the system, as the following example shows: "*I received a parental message, I do not know if it was the older brother or the parent who sent the message.../... so I needed to go back to the paper notepad to write a note.../... on the notepad, there's the handwriting, the signature, we can quickly tell the difference between a parent and a child*".

### 5.4 Factors Linked to Social and Work-related Organization

VLEs were particularly appreciated as a tool supporting communication (72 positive comments). Certain teachers, who created blogs, mentioned these blogs in the notepads when information needed to be consulted. Teachers seemed to appreciate the positive role that the VLE played in teacher collaboration (12 mentions). Sharing resources made it easier to organize common activities and outings, and facilitated pedagogical work.

Negative comments (8) addressed the messaging service as a communication method, highlighting the fact that this service did not distinguish between in-school time and out-of-school time. Teachers mentioned the need to change the settings so that parents could only send messages outside of school time and to limit school-time messages between students. Concerning the parents, such parameters would limit the amount of last-minute intrusive messages that require additional work on the teacher's behalf during class time. Teachers have more control using the parent-teacher notepad. Providing these settings could be useful as a first step. It would reassure teachers and would give them time to set-up digital awareness activities for students and parents.

### 5.5 Factors linked to the Tool's Use and Deployment

Teachers reported finding the platform user-friendly (27 positive comments). They considered the functionalities and information coherent and easily accessible through the menu and the icons. The negative comments (24) were linked to the functionalities in the VLE's Document space: teachers would have liked to share folders rather than files: *"the children receive... [the files] just like that. It is not easy for them, we have a Shared Document and everything is mixed together: music, stories. If the name of the folder is a bit vague, they will not know"*. There was also a lack of visibility as to who consulted content and who connected to the platform. By following the news feed, teachers managed to see the activity of other users (parents, students), but only if the latter had modified a certain feature, such as their avatar or their motto. But feedback could not be retrieved when users simply consulted the platform without leaving tangible traces. *"It is true that... if they do not change their mood or their motto, we do not know if they have connected or not. It would be interesting for us users to know who saw the content"*. In order to obtain such data, teachers had to do an additional task which consisted in sending a questionnaire through the parent-teacher notepad or by asking the students if their parents had connected to the platform. Such feedback was important in order to build ties with the different educational partners and to make sure that the published information had actually been seen and received. Otherwise, teachers had difficulties knowing if the system was really useful and effective.

The lack of computer infrastructure (equipment, networks...) was also seen as hampering the acceptance of VLEs (6 comments). Teachers would have liked to use the VLE in class with the students but they did not have enough computers and tablets. *"we would almost need to have computers in the class all the time to really use (VLEs) in every day teaching"*. Teachers also pointed out that all students did not have equal access to VLEs: some had continuous access, while others had restricted access through their parents; some students did not have Internet access at all. Finally, teachers mentioned a lack of support and assistance. They did not feel adequately trained to use VLEs. Given the fact that this was an experimental implementation phase, not all possible means were used to support the teachers. On the long term, academic supervisors would need to get involved in training and supporting teachers.

## 6   Discussion and Conclusion

In this paper we presented three important models in the study of technological adoption. The three models have their origins in different fields of research. The models of "social acceptance", like TAM and UTAUT were inspired by social psychology but applied to management and marketing studies. The "practical acceptance" theories are specific to ergonomists and designers. And finally, models of "situated acceptance" are also issued from development psychology and lately applied to various fields, from change management to organization issues. The technological adoption issue is of general interest and should not be limited to one singular approach. Our objective was to resume these various models and to extract information that is salient for educational area. Factors like usability for young children, teachers' confidence in their computer use skills, teacher's perceived security toward children's use and preexistent teaching practices are example of important determinants of technology adoption in schools.

The study of teachers' acceptance done according to the model shows that, in terms of acceptance, the uses of the VLE One spurred tensions that were similar to the ones described by Prieur and Steck [4] and Voulgre [11] in secondary education. We observed contradictions between the artefact, the community and the rules as well as contradictions between the artefact and the division of work. The first type of contradiction was linked to the subverted uses of the Messaging Service or the News Feed. There was also a lack of digital access due to poor infrastructure in schools and in some homes. The second type of contradiction was due to an excessive workload and an increase in the teachers' professional responsibilities through the extension of the "school space-time continuum". We recommend that decision-makers (the Ministry, school districts) provide better information on VLE users' responsibilities. When it comes to community uses – such as the ways in which to use the messaging service or whether or not use feedback indicators – we think that such decisions can be made at a local level through discussions between the school administration, the teachers and the VLE publisher. Depending on contexts and practices, certain modes of operation may or may not be effective or acceptable.

There were fewer contradictions linked to the artefact itself. Teachers appreciated the services offered by One as well as its ergonomics; they tried to adapt the VLE to their professional practices. They did not hesitate to make requests to improve the tool. They also agreed to help train children and their parents on digital best practices. Teachers showed signs of acceptance in this area, but they still need to be given more support and assistance to maintain such uses on the long term.

To conclude, the acceptance of this VLE seems to have been overall positive since One was well designed and relatively adapted to the practices of the teachers involved. The main problems are linked to the ways in which the tool is implemented. The recommendations formulated here are meant for the Ministry of Education and school principals. Clarifications need to be made concerning the limits of the school space-time continuum and the rules of governance and communication. Such resolutions are relevant in a context in which very young children are concerned, since they are to use these platforms without having prior social digital skills.